\documentclass[twocolumn]{aastex63}

\shorttitle{QPO of Cygnus X-1 in the soft state}
\shortauthors{Yan et al.}
\graphicspath{{./}{figures/}}

\begin{document}

\title{Detection of a low frequency quasi-periodic oscillation in the soft state of Cygnus X-1 with Insight-HXMT}

\correspondingauthor{Zhen Yan}
\email{zyan@shao.ac.cn}

\author[0000-0002-5385-9586]{Zhen Yan}
\affiliation{Shanghai Astronomical Observatory, Chinese Academy of Sciences, \\ 80 Nandan Road, Shanghai 200030, China }


\author{Stefano Rapisarda}
\affiliation{Shanghai Astronomical Observatory, Chinese Academy of Sciences, \\ 80 Nandan Road, Shanghai 200030, China }

\author{Wenfei Yu}
\affiliation{Shanghai Astronomical Observatory, Chinese Academy of Sciences, \\ 80 Nandan Road, Shanghai 200030, China }




\begin{abstract}
 We report the detection of a short-lived narrow quasi-periodic oscillation (QPO) at $\sim$88 mHz in an Insight-HXMT observation during the soft state of the persistent black hole high-mass X-ray binary Cygnus X-1. This QPO is significantly detected in all the three instruments of Insight-HXMT, so in the broad energy range 1-250 keV. The fractional rms of the QPO does not show significant variations above 3 keV ($\sim$5\%) while it decreases at lower energy ($\sim$2\%). We show that this QPO is different from the type-A, -B, and -C QPOs usually observed in black hole X-ray binaries. We compare QPOs at similar frequencies that have been previously detected in other persistent high mass X-ray binaries in the soft state; we speculate that such QPOs might relate to some local inhomogeneity rarely formed in the accretion flow of wind-fed accretion systems.

\end{abstract}



\section{Introduction} \label{sec:intro}
Cygnus X-1 is a bright persistent X-ray binary consisting of a black hole accreting matter from an OB star \citep{Bowyer1965,bolton_identification_1972,webster_cygnus_1972}. Since its discovery, Cygnus X-1 has been the target of extensive multiwavelength observational campaigns \citep[e.g. ][]{herrero_fundamental_1995, nowak_rossi_1999, gallo_dark_2005, albert_very_2007,grinberg_long_2013, kantzas_new_2021}, making it one of the most-studied X-ray sources. From the very early observations of Cygnus X-1, it was clear that the X-ray emission undergoes dramatic changes in spectral distribution and brightness \citep{tananbaum_observation_1972}.

 Analogous changes have been observed in most later discovered black hole X-ray binaries (BH XRBs) and have been classified into different accretion states  \citep[e.g. ][]{remillard_x-ray_2006, belloni_states_2010}. The accretion states that have been first identified are the low-hard state (LHS) and the high-soft state (HSS). In the LHS, the energy spectrum can be described by a power law (the photon index $\sim$1.5--2.1) and in the HSS by a multitemperature blackbody component. It is generally accepted that the multitemperature blackbody emission arises from an accretion disk around the black hole \citep{shakura_black_1973} and the power law is the result of Compton upscattering of cool photons by hot electrons close to the black hole \citep[see reviews in ][]{done_modelling_2007,yuan_hot_2014}. Accretion states are also identified according to the timing properties of the source \citep[see reviews by ][]{remillard_x-ray_2006,belloni_states_2010,belloni_fast_2014}. The power density spectrum (PDS) in the LHS usually exhibits low-frequency quasi-periodic oscillations (LF QPOs, $\sim$ 0.01--10 Hz) on top of broadband noise. QPOs are classified as type A, B, and C according to their frequency, amplitude, and coherence \citep[see ][ for detailed definitions of QPO types]{wijnands_complex_1999, casella_study_2004}. Strong high-coherence type-C QPOs are usually observed in the LHS, when the total fractional rms amplitude is high ($\sim$30\%). In the HSS the amplitude of the variability drops dramatically ($ \lesssim$1\%) and no QPOs are observed. The above phenomenon has been observed in almost every transient BH XRB, which has the largest population of known Galactic BHs. The LHS is usually observed at the beginning and the end of periods of enhanced accretion (outbursts), when the luminosity is low, while the HSS occurs at maximum luminosity. Between these two extremes, several transition states are also observed \citep{remillard_x-ray_2006,belloni_states_2010}.

 However, some persistent BH XRBs can exhibit accretion states different from the transient BH XRBs. Cygnus X-1 is one of these sources, as its HSS is characterized by both multitemperature blackbody and Compton power law emission \citep[e.g. ][]{gierlinski_radiation_1999}. The X-ray timing properties of accretion states in Cygnus X-1 differ from the general description mentioned above. Type-C QPOs have never been detected during the LHS of Cygnus X-1 \citep{grinberg_long_2014,ingram_review_2019} and the HSS shows high variability \citep[fractional rms amplitude $\sim$25\%;][]{axelsson_evolution_2005,grinberg_long_2014}. However, there were broad and weak narrow QPOs detected in its PDS
 during different spectral states, which were sometimes called QPOs \citep[e.g., ][]{cui_temporal_1997,paul_low_1998,pottschmidt_long_2003,axelsson_evolution_2005,shaposhnikov_comprehensive_2006,grinberg_long_2014}.

 In this paper, we report the detection of a millihertz QPO in one of the Insight-HXMT observations of Cygnus X-1 during the soft state. In Section \ref{sec:obs}, we describe the data reduction, in Section \ref{sec:data_an} the spectral and timing analysis, and in Section \ref{sec:disc} we compare the detected signal to similar signals observed in other BH XRBs and discuss possible physical scenarios. 

\section{Observations and Data Reduction}
\label{sec:obs}
 \begin{figure*}
\centering
	\includegraphics[width=\textwidth]{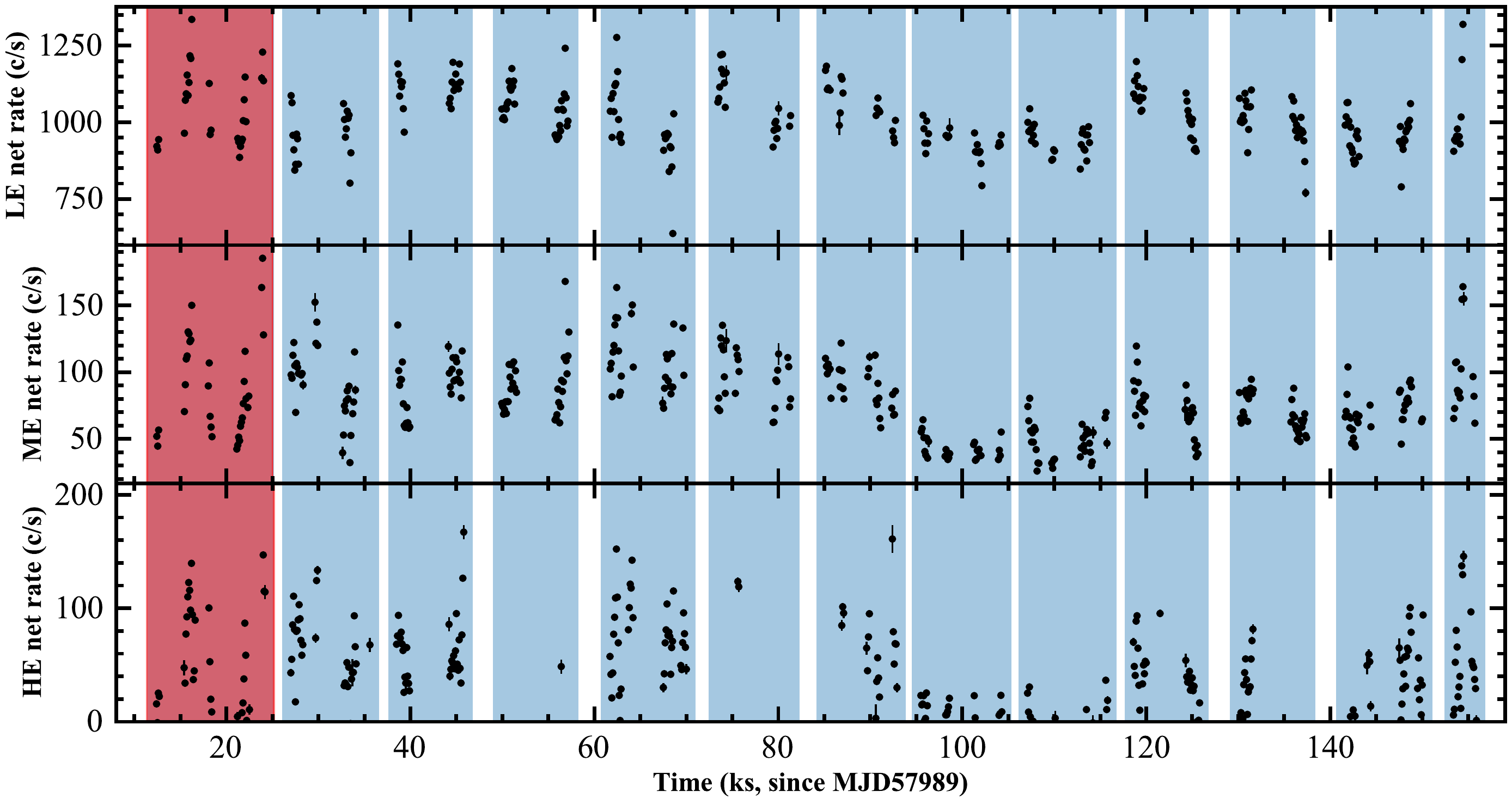}
    \caption{The light curves with a time resolution of 100 s during the Insight-HXMT observations for the three independent payloads: LE, ME and HE. Each exposure is marked by the shaded region. The red one is the exposure during which the QPO is detected. }
    \label{fig:lc}
\end{figure*}

The Insight-Hard X-ray Modulation Telescope (HXMT) \citep{zhang_overview_2020} is the China's first X-ray astronomy satellite. Insight-HXMT has three major scientific payloads: the Low Energy Telescope (LE, ~1--10 keV), the Medium Energy Telescope (ME, ~5--35 keV), and the High Energy Telescope (HE, ~20-250 keV), capable of  observations down to a time resolution of 1 ms, 276$\mu$s, and 25$\mu$s, respectively. The effective area of the three payloads is 384 cm$^{2}$,  952 cm$^{2}$, and 5000 cm$^{2}$, respectively.

In order to reduce the data file size, a long Insight-HXMT observation is man made, split into multiple segments. Each segment lasts approximately three hours, which is named as ``exposure" and identified with an Exposure ID\footnote{\url{http://hxmtweb.ihep.ac.cn/SoftDoc/67.jhtml}}. We analyzed the longest observation of Cygnus X-1 (Obs. ID P0101315001, start time 2017 August 24T02:49:21 UTC) in the Insight-HXMT data archive. The selected observation contains 13 exposures, for a total observation time of about 150 ks (see \autoref{fig:lc}).

We performed data reduction by using the Insight-HXMT Data Analysis Software package (\texttt{HXMTDAS}) V2.02. For all the three payloads, we extracted cleaned events applying the following filtering criteria: (1) exclusion of time periods when Insight-HXMT passes through the South Atlantic Anomaly (T\_SAA $>$ 300 \&\& TN\_SAA $>$ 300 \&\& SAA\_FLAG == 0); (2) elevation angle $>$ 10$^\circ$ (ELV $>$ 10); (3) pointing offset angle $<$ 0.1$^{\circ}$ (ANG\_DIST $<$ 0.1); (4) cutoff rigidity $>$ 8 GeV (COR $>$ 8). An additional filtering criterion was applied for LE data: elevation angle from Earth bright limb $>$ 20$^{\circ}$ (DYE\_ELV $>$ 20$^{\circ}$). For LE, ME, and HE, we computed binned light curves, energy spectra, and corresponding background from cleaned event files using the \texttt{HXMTDAS} tools $<$le/me/he$>$lcgen, $<$le/me/he$>$specgen, and $<$le/me/he$>$bkgmap, respectively. In particular, we extracted source and background light curves with time resolution 1/256 ($\approx$0.0039) s in the energy band 1--10 keV, 5--30 keV, and 20--250 keV  for LE, ME, and HE, respectively.

\begin{figure*}
\centering
	\includegraphics[width=\textwidth]{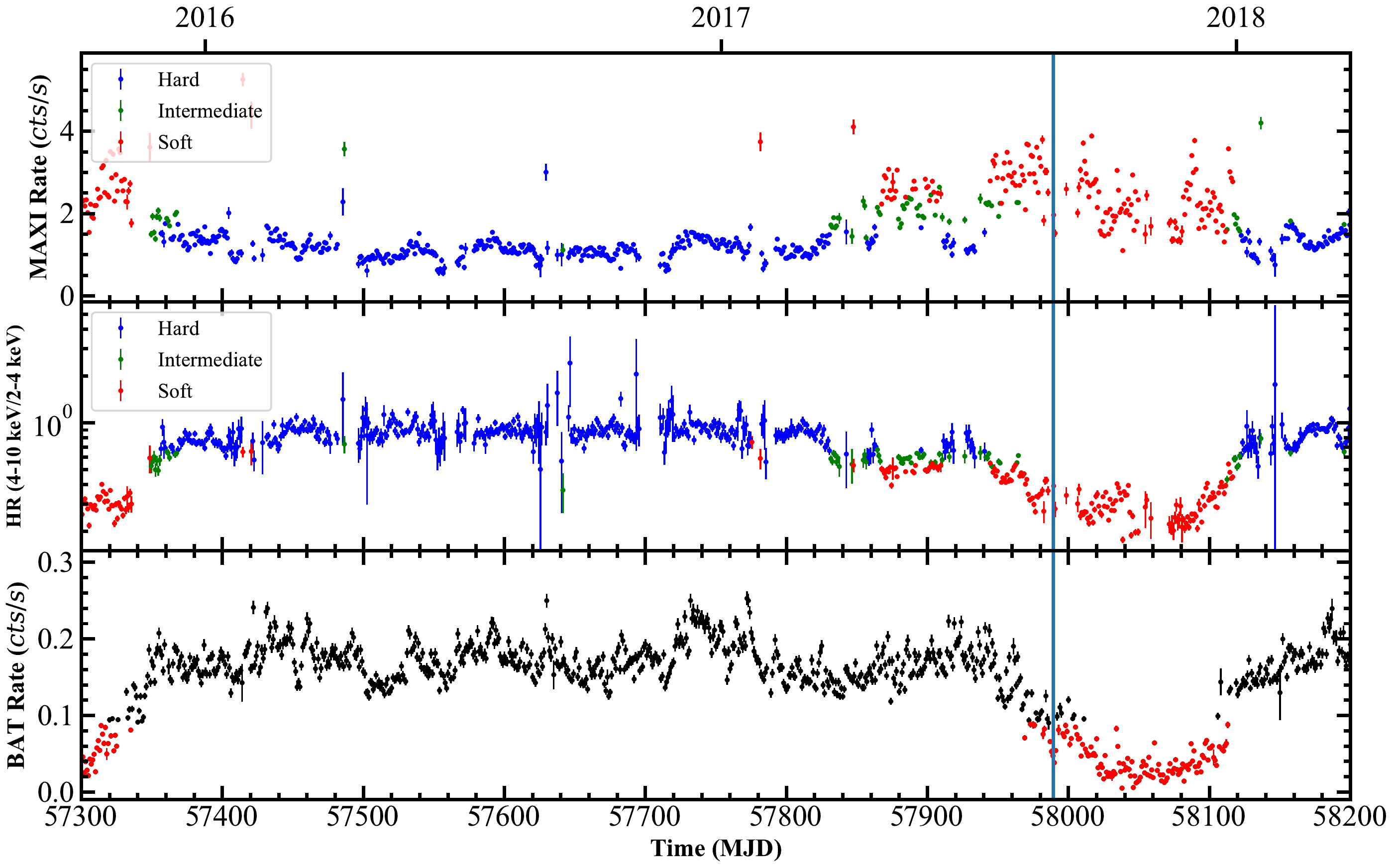}
    \caption{The daily light curves of Cygnus X-1 obtained from MAXI and Swift/BAT monitoring products. The definitions of X-ray spectral states are adopted from \citet{grinberg_long_2013}. The vertical solid line marks the time of the  Insight-HXMT observation we reported here.
    }
    \label{fig:maxi-bat}
\end{figure*}

We also downloaded the daily light curves from the X-ray all-sky monitoring instruments Swift/BAT \citep{krimm_swiftbat_2013} and MAXI \citep{matsuoka_maxi_2009}, selecting Swift/BAT and MAXI data simultaneous to our Insight-HXMT observation (MJD 57989; see \autoref{fig:maxi-bat}).

\section{Data Analysis and Results}
\label{sec:data_an}

\subsection{Timing Analysis}
\label{sec:timing}
  \begin{figure*}
\centering
	\includegraphics[width=\textwidth]{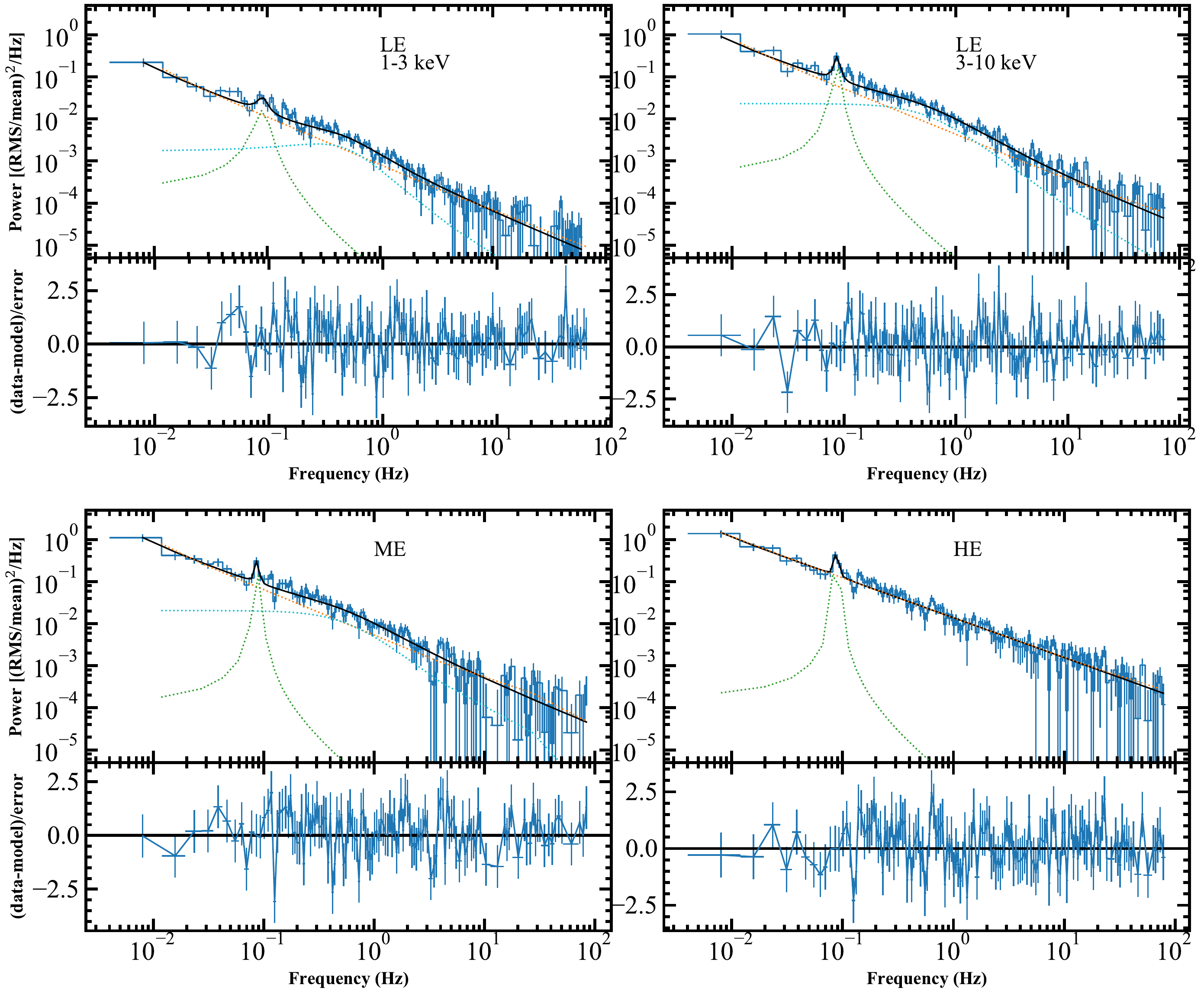}
    \caption{ The PDSs of Cygnus X-1 of the LE, ME and HE on board the Insight-HXMT satellite. The solid black line is the best-fitting model of the PDS of each payload. The orange and green dotted lines show the power law component and the QPO. The cyan dotted line in the LE and ME subfigures is the BLN component. The lower panel of each subfigure is the ratio between residual and the error of the data. }
    \label{fig:pds}
\end{figure*}
 For each payload and exposure, we subtracted the background from the light curve, we divided the light curves into segments of 128 s, and, for each segment, we computed Leahy-normalized PDSs \citep{leahy_searches_1983} using the \texttt{HEASARC} tool \texttt{powerspec} (XRONOS package). PDSs were averaged, obtaining in this way a single PDS for each exposure. Because of the chosen time resolution and segment length, PDSs have frequency resolution $d\nu = 1/128 = 0.0078125$ Hz and Nyquist frequency $\nu_{N} = 128$Hz. 
 
\autoref{fig:lc} shows the light curve with a time resolution of  100 s for each payload for the entire selected observation. We visually inspected PDSs for each exposure and find obvious QPO features in all three payloads during the first exposure (Exp. ID P010131500101-20170824-01-01; red region in \autoref{fig:lc}). In this exposure, the PDS of each payload is characterized by a power law broadband noise with a narrow feature on top of it at low frequency (\autoref{fig:pds}). This narrow feature appears to be more prominent at higher energies (ME and HE), and it is not present in any other exposures of the selected observation. To ascertain the presence of the narrow feature in the LE low-energy regime (1--10 kev), we performed the same data reduction and analysis described above in two sub-low-energy bands, 1--3 keV and 3--10 keV, respectively.

We then subtracted the contribution from the Poisson noise, which was estimated by averaging Leahy-normalized PDSs above 100 Hz. The noise levels (1.99, 2.00, 1.99, 2.09 for LE 1--3 keV, LE 3--10 keV, ME, and HE) are consistent with the expected Poisson noise level in Leahy-normalized PDSs  \citep{leahy_searches_1983}. The average net count rate 870.50,178.00, 87.04, and 52.60 counts/s for LE(1--3 keV), LE(3--10 keV), ME, and HE are used for calculating rms-normalized PDSs \citep{BelloniHasinger1990}. The total fractional rms amplitude of each PSD is 11.4\%, 27.3\%, 28.7\%, and 37.1\% (below 32 Hz). We then fit the PDSs using a model consisting of a power law plus a narrow Lorentzian. The fitting was performed in \texttt{XSPEC}  \citep{arnaud_xspec_1996}. We obtained a statistically acceptable fit for the HE PDS (see \autoref{fig:pds} and \autoref{tab:pds_fit}), while we noticed broad frequency residuals around 0.5 Hz for the LE and ME PDS. Adding a band-limited noise (BLN; a Lorentzian function with $\nu_{0}$ fixed at 0) in the model, we obtain a statistically acceptable fit also for LE and ME (see \autoref{fig:pds} and \autoref{tab:pds_fit}). We then calculate the $\nu_\mathrm{max}$ ($=\sqrt{\nu_{0}^{2}+(FWHM/2)^{2}}$) to represent its characteristic frequency \citep[e.g.][]{belloni_unified_2002}, the values of which are $0.50^{+0.07}_{-0.07}$, $0.56^{+0.08}_{-0.07}$ and $0.56^{+0.13}_{-0.11}$ for LE(1--3 keV), LE(3--10keV), and ME data.  

\begin{figure}
\centering
	\includegraphics[width=\linewidth]{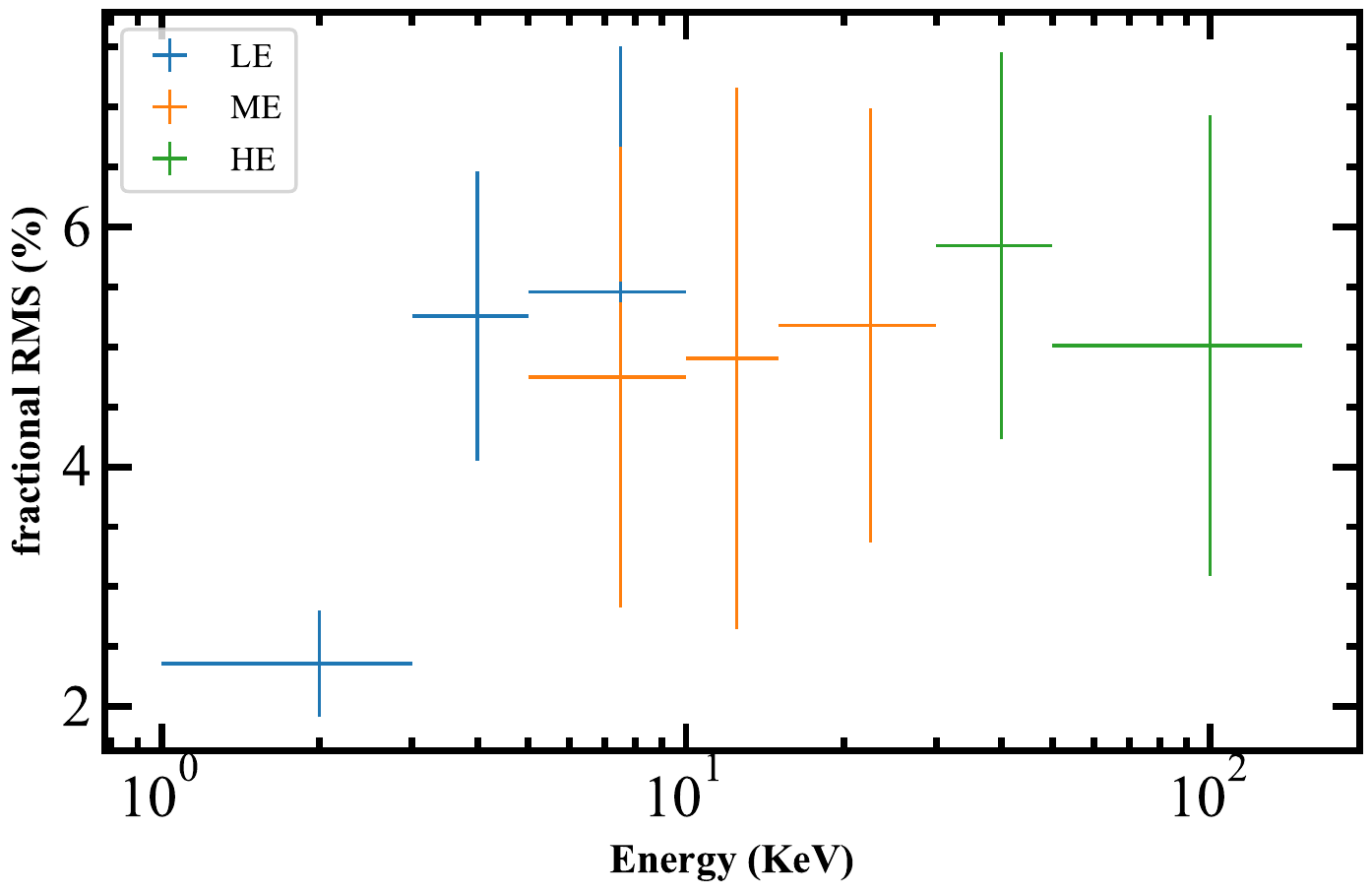}
    \caption{The fractional rms of the QPO as a function of photon energy. It becomes almost constant ($\sim$5\%) above 3 keV and extends to photon energies beyond 50 keV.}
    \label{fig:rms}
\end{figure}

All the best-fitting parameters are listed in \autoref{tab:pds_fit}. The power-law noise with an index of $\sim$1 and the fractional rms are consistent with the soft state of Cygnus X-1 \citep[e.g. ][]{cui_rossi_1997,axelsson_evolution_2005,axelsson_probing_2006,grinberg_long_2014}. A narrow QPO is detected in the all PDSs with significance of 4.7$\sigma$, 6.8$\sigma$, 5.6$\sigma$, and 6.3$\sigma$ for the LE(1--3keV), LE(3--10keV), ME, and HE, respectively. The QPO frequencies at different energy bands are consistent within uncertainties; the averaged value is $\sim$88 mHz. The QPO is very narrow, especially at higher-energy bands (see the FWHM values in \autoref{tab:pds_fit}). The quality factors ($Q=\nu_{0}/FWHM$) are $10.8\pm7.7$, $15.3\pm8.1$, $>13.7$, and $>13$ for the LE(1--3keV), LE(3--10keV), ME, and HE, respectively. We further investigated the energy dependence of the QPO rms with more subenergy bands. We first produced the PDSs in the energy bands of 1--3 keV, 3--5 keV, 5--10 keV for LE data; 5--10 keV, 10--15 keV, 15--30 keV for ME data; and 30--50keV, 50--150 keV for HE data. We then calculated the fractional rms of the QPO by fitting each PDS. The QPO rms becomes almost constant ($\sim$ 5 \%) above 3 keV and up to at least 50 keV (see \autoref{fig:rms}).

\begin{table*}
\centering
\caption{Best-fitting Parameters of the LE and ME PDSs}
\label{tab:pds_fit}
\begin{tabular}{lcccccc}
\hline
\hline
Component & Parameter  & LE    &  LE  & ME  & HE\\
           &            & (1--3keV) & (3--10keV) &  \\
\hline
POW        &Index      & $1.16^{+0.04}_{-0.04}$ & $1.08^{+0.03}_{-0.03}$ &     $1.08^{+0.04}_{-0.04}$    &$0.95^{+0.02}_{-0.02}$    \\
\hline
LOR        & $\nu_{0}$ (mHz) &    $89.27^{+5.68}_{-2.87}$  &  $86.91^{+1.47}_{-1.71}$   &   $87.53^{+1.70}_{-2.63}$ &$88.84^{+0.72}_{-1.48}$\\
(QPO)      & FWHM  (mHz)   & $14.03^{+15.53}_{-9.35}$   &   $7.39^{+4.86}_{-3.75}$    &      $2.88^{+3.62}_{-2.88}$  & $2.64^{+4.19}_{-2.64}$ \\
           & rms(\%)    &     $2.18^{+0.44}_{-0.47}$     &  $5.82^{+0.76}_{-0.87}$  & $4.75^{+0.79}_{-0.91}$  &$5.94^{+0.89}_{-0.94}$\\
\hline
LOR       &$\nu_{0}$ (Hz)  &0 & 0   &0   & \nodata  \\
(BLN)     & FWHM (Hz)    &$0.99^{+0.15}_{-0.13}$ & $1.12^{+0.15}_{-0.13}$ & $1.11^{+0.26}_{-0.22}$  & \nodata \\
         &rms(\%)    & $5.57^{+0.39}_{-0.41}$          & $14.31^{+0.74}_{-0.77}$    &    $13.39^{+1.12}_{-1.18}$  & \nodata \\
\hline

        &$\chi^{2}$/dof    &  161.08/156  &  189.59/156 & 163.90/156   & 178.93/158\\
\hline

\hline 
\end{tabular}   
\tablenotetext{}{
Note: The model is \texttt{Powerlaw+Lorentzian+Lorentzian} for LE and ME, \texttt{Powerlaw+Lorentzian} for HE. The uncertainties are the 1$\sigma$ ranges.}
\end{table*}

\subsection{Further assessing the significance of the QPO}
  \begin{figure*}
\centering
	\includegraphics[width=\textwidth]{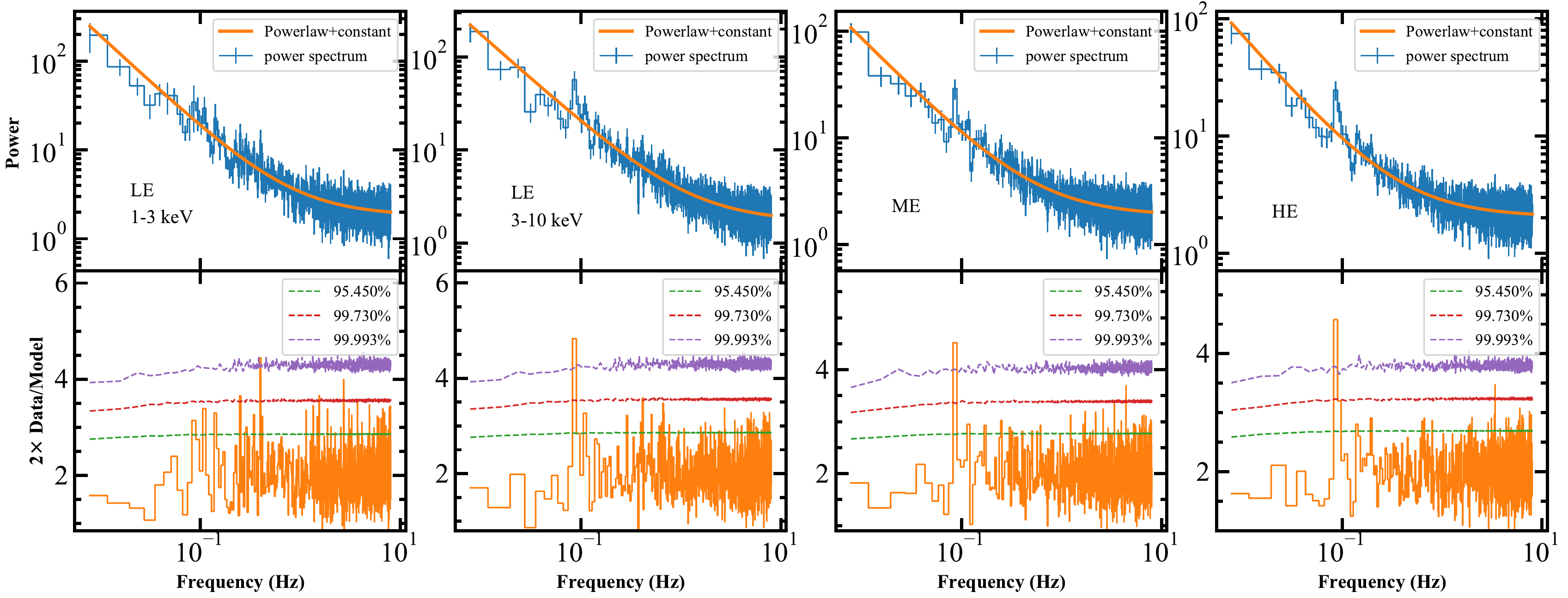}
    \caption{Upper panel: The PDS of LE (1--3keV),LE (3--10 keV), ME and HE. The orange line marks the best fitting model composed of a power law plus a constant.  Lower panel: the ratio of the PDS data to the best-fitting power law model (the orange line in the upper panel). The green, red and purple dashed lines mark the levels at 95.450\%,99.730\% and 99.993\% of the simulated PDSs.}
    \label{fig:percentile}
\end{figure*}

 The classical method to determine the detection level of a signal is based on the statistical properties of the white noise \citep{van_der_klis_fourier_1989}. For an averaged PDS, the white-noise power follows a $\chi^{2}$ distribution with $2MW$ dof, where $M$ is the number of averaged powers at a specific frequency and $W$ is the number of averaged frequency bins. Integrating the $\chi^{2}$ probability density function, it is possible to determine the $p$-value of a given power and define detection levels above the underlying noise. However, this method is not reliable for accessing detection above noise processes differing from white noise \citep{israel_new_1996,vaughan_simple_2005,vaughan_bayesian_2010}. 
 
 In our case, the QPO is detected in a frequency range dominated by red noise. To access the significance of the QPO with respect to the red noise, we applied the method proposed by \citet{vaughan_bayesian_2010} and implemented in the Python package \texttt{stingray} \citep{huppenkothen_stingray_2019}. We only considered the PDS below 8 Hz, where the red noise dominates. We fitted the PDSs with a power-law function plus a constant and sample the model parameters with a  Markov Chain Monte Carlo (MCMC) method in the \texttt{emcee} package \citep{foreman-mackey_emcee:_2013}. We then generated $10^{5}$ fake PDSs using the sampled parameters, in order to derive a posterior probability distribution for the test statistics and thus a posterior predictive $p$-value. Our observed PDSs are averaged by $M$ 128s segments, where $M$ is 18, 22, and 27 for LE, ME, and HE. We considered the $M$ when generating the fake PDSs and did not rebin the fake PDSs. So, the derived $p$-value has been corrected for the number of periodograms searched including the number of segments and the number of frequencies.

We fitted each simulated PDS with a power-law plus a constant model and calculate $(2I_{j}/S_{j})$, where $I_{j}$ is the power at a given frequency $f_{j}$ of a PDS and the $S_{j}$ is the power from the model. We then plot the 99.450\%, 99.730\%, and 99.993\% (2$\sigma$, 3$\sigma$ and 4$\sigma$) significance levels, which is obtained from the  $10^{5}$  samples of the $(2I_{j}/S_{j})$ at each frequency (\autoref{fig:percentile}). There is only one observed power at the frequency $\sim$86 mHz is larger than that of the 99.993\% simulated PDS, which indicates a significant signal ($>4\sigma$) around this frequency in the PDS of LE(3--10keV), ME and HE. In the PDS of LE(1--3keV), the power at $\sim$ 86 mHz is marginally at $2\sigma$. 

\begin{figure*}
\centering
	\includegraphics[width=\textwidth]{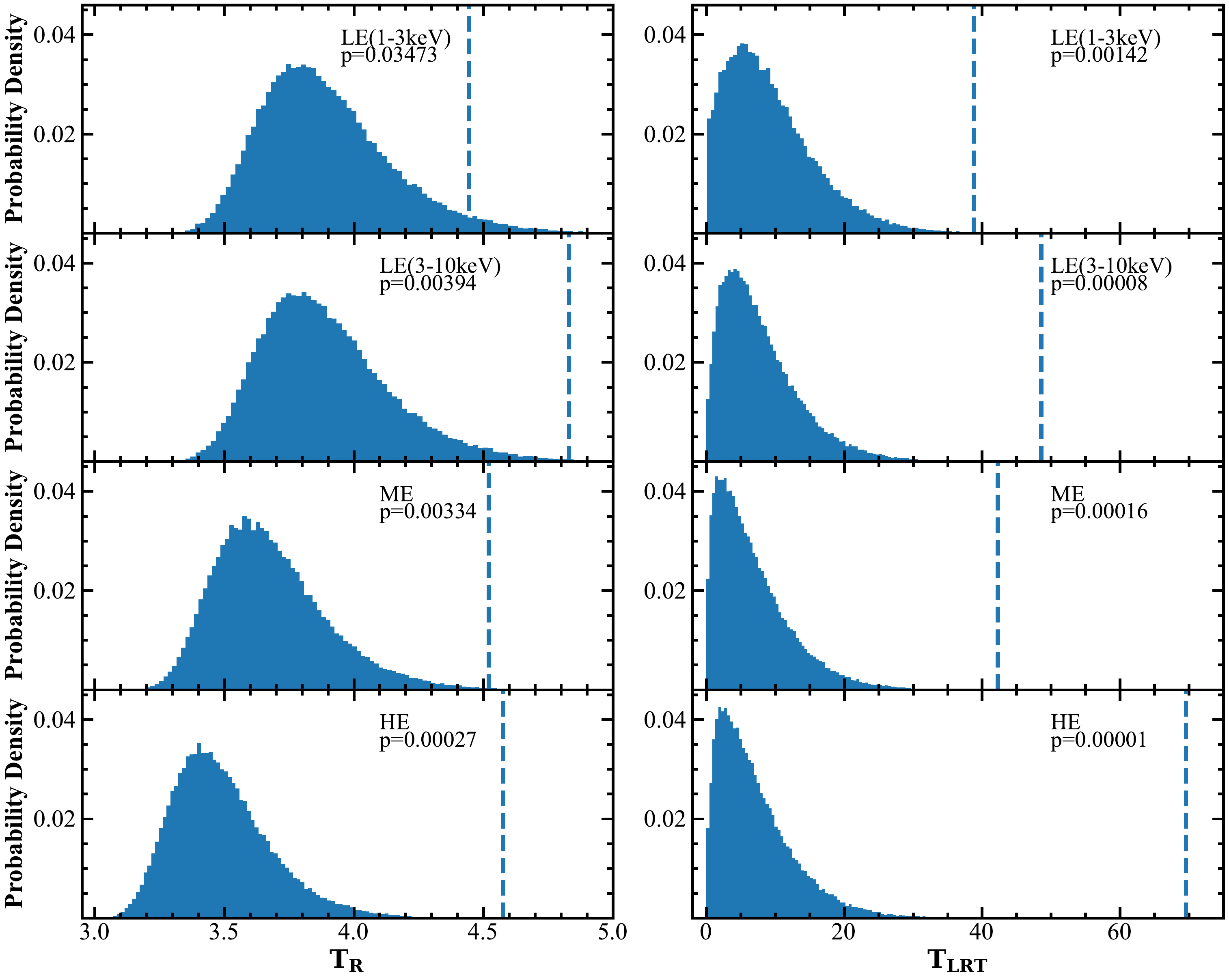}
    \caption{The posterior probability distributions of the test statistics $T_\mathrm{R}$ and $T_\mathrm{LRT}$ from the simulated sample. The dashed line marks the observed value of $T_\mathrm{R}$ and $T_\mathrm{LRT}$. The posterior predictive $p$-value is the probability that the simulated values larger than the observed one.}
    \label{fig:pvalue}
\end{figure*}

We first defined the test statistic $T_\mathrm{R}=max(2I_{j}/S_{j})$. Then the $p$-values of the observed $T^{obs}_\mathrm{R}$ from the posterior probability distribution of $T_\mathrm{R}$  are 0.00394, 0.00334, and 0.00027 for LE (3--10 keV), ME, and HE, respectively \autoref{fig:pvalue}. The combined $p$-value of the three is 7.45$\times 10^{-7}$ by using the Fisher method since they are independent tests for the three instruments. The frequency with maximum $T_\mathrm{R}$ from the three instruments are the same, so the $p$-value at this specific frequency should be smaller than the combined $p$-value \citep[e.g. ][]{huppenkothen_detection_2017}, demonstrating a very significant signal of $\sim86$ mHz existing above the red noise. 

We also used likelihood ratio test (LRT) to compare the two models with/and without a Lorentzian component. The likelihood ratio is defined as $T_\mathrm{LRT}=-2\log(L_{0}/L_{1})$, where the $L_{0}$ is the likelihood of the simple model and the $L_{1}$ is the likelihood of the complex model. We fitted each simulated PDS with the two models and computed the $T_\mathrm{LRT}$. The posterior probability distributions of the $T_\mathrm{LRT}$ can give us the $p$-values of $T^{obs}_\mathrm{LRT}$ are 0.00008,0.00016, and 0.00001 for LE (3--10 keV), ME, and HE respectively \autoref{fig:pvalue}. Therefore, the statistics of $T_\mathrm{R}$ and $T_\mathrm{LRT}$ both demonstrate that a significant QPO detection in LE(3--10 keV), ME, and HE. The $p$-values of $T_\mathrm{R}$ and $T_\mathrm{LRT}$ of LE (1--3 keV) are at least one order of magnitude larger than those of higher-energy bands, which means the QPO is less significant below 3 keV. 

\subsection{Spectral  and color analysis}
\begin{figure*}
\centering
	\includegraphics[width=\textwidth]{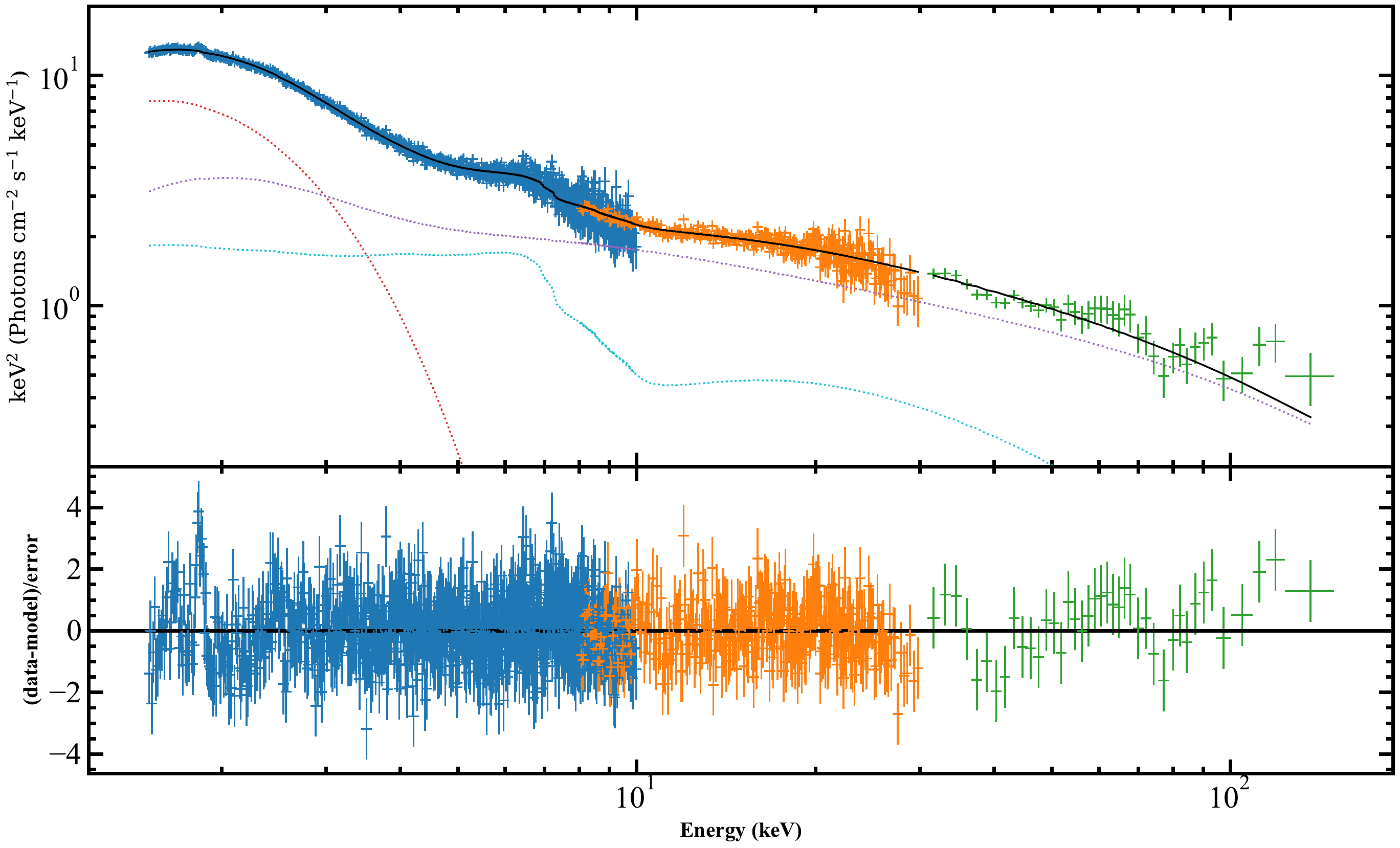}
    \caption{Upper panel: the unfolded spectrum of the LE (blue), ME(orange), and HE (green) and the model \texttt{const*tbabs*(diskbb+reflkerrG)} (black solid line). The dotted red, purple, and cyan lines represent the disk, Comptonization, and reflection components. Lower panel: ratio of the residual to the error of the data.}
    \label{fig:SED}
\end{figure*}

  We performed spectral analysis of the selected observation by combining the data from LE, ME, and HE. The effective areas of the three instruments have been calibrated using Crab nebula observations as described in \cite{li_-flight_2020}. We jointly fit energy spectra of LE (1--10 keV), ME (8--30 keV), and HE (30--150 keV) with the model \texttt{const*tbabs*(diskbb+reflkerr)} (see \autoref{fig:SED}), where abundances and cross sections of the absorption by the Galactic interstellar medium are set according to \citet{wilms_absorption_2000} and \citet{verner_atomic_1996}. \texttt{reflkerr} is a relativistic reflection model that computes direct and reflection spectra assuming either a slab or a spherical plasma geometry \citep{niedzwiecki_improved_2019}. In this framework, the Comptonization spectrum is computed with \texttt{compps} \citep{poutanen_two-phase_1996}. We fix the black hole spin \citep[0.99, ][]{zhao_confirming_2020,zhao_re-estimating_2021}, the inclination angle \citep[27$^{\circ}$.5,][]{miller-jones_cygnus_2021},  and the index of the outer disk $q_\mathrm{out}$ (3), and we leave the inner disk index $q_\mathrm{in}$ and break radius $R_\mathrm{br}$ free \citep{fabian_determination_2012,walton_soft_2016}. We also tied the temperature of seed photons for Comptonization with the disk temperature. All of the best-fitting parameters are listed in \autoref{tab:spec_fit}. According to the electron temperature $kT_\mathrm{e}$ and  optical depth $\tau$ of the corona, we can derive the photon index of the Comptonization component \citep{zdziarski_broad-band_1996}.  The derived photon index $\Gamma=2.74\pm0.11$ and the best-fitting disk temperature $T_\mathrm{in}=0.49\pm0.01$ keV are consistent with the soft state of Cygnus X-1 \citep{tomsick_reflection_2014, walton_soft_2016,kawano_black_2017,lubinski_distinct_2020}. The unabsorbed X-ray flux (0.1-100 keV) is 8.16$\times$10$^{-8}$ ergs s$^{-1}$ cm$^{-2}$ , which corresponds to $\sim$2\% Eddington luminosity at a distance of 2.22 kpc and a BH mass 21.2 $M_{\odot}$ \citep{miller-jones_cygnus_2021}.

To confirm that Cygnus X-1 is in the soft state in our selected exposure, we computed count rates from swift/BAT and MAXI light curves simultaneous to our data (MJD 57989; see \autoref{fig:maxi-bat}). The count rate in the 15--50 keV (BAT) and 2--4 keV (MAXI) energy band is 4.73$\pm0.28 \times10^{-2}$ counts cm$^{-2}$ s$^{-1}$ and 1.97$\pm$0.23 counts s$^{-1}$, respectively. The corresponding hardness ratio (4--10 keV/2--4 keV) is 0.39$\pm$0.08. According to the state classification criteria of Cygnus X-1 described in \citet{grinberg_long_2013}, the observed X-ray intensity and hardness ratio are both consistent with the source being in the soft state.

\begin{table}
\centering
\caption{Best-fitting parameters of the energy spectrum}
\label{tab:spec_fit}
\begin{tabular}{lccc }
\hline
\hline
Component  & Parameter &  & Value \\
\hline
TBABS &     $n_\mathrm{H}$  &  [$10^{22}~cm^{-1}$] & $0.46^{+0.04}_{-0.04}$\\
DISKBB&    $kT_\mathrm{in}$ &  [keV] &0.49$^{+0.01}_{-0.01}$  \\
       &  Norm &  [$10^{4}$]  &  $2.64^{+0.17}_{-0.19}$  \\
REFLKERR & $q_\mathrm{in}$ &  &$7.52^{+0.58}_{-1.75}$ \\
          &$R_\mathrm{br}$  & [$r_{g}$] &   $3.29^{+0.23}_{-0.21}$  \\
          &$R_\mathrm{in}$ & [$r_{g}]$ &$2.22^{+0.10}_{-0.16}$ \\
          &$\tau$ & &$0.59^{+0.04}_{-0.03}$  \\
          &$A_\mathrm{fe}$ & [solar] & $<10$ \\
          &$kT_{e}$ &[keV] & $87.74^{+2.82}_{-2.39}$ \\
          &$\log\xi$ &log[erg cm  s$^{-1}$] & $4.33^{+0.06}_{-0.03}$ \\
          & $f_\mathrm{refl}$ & & $0.44^{+0.08}_{-0.07}$\\
\hline
$\chi^{2}/dof$  & & & 1332.78/1445 \\
\hline 

\end{tabular}
\tablenotetext{}{
Note: The uncertainties are the 1$\sigma$ ranges.}
\end{table}

\section{Discussion}
\label{sec:disc}
We detected a very low-frequency ($\sim 88$ mHz) QPO only during the first 12 ks of the longest Insight-HXMT observation of Cygnus X-1. This QPO is significantly detected in LE, ME, and HE, independently. The fractional rms amplitude is energy independent (above 3 keV; see \autoref{fig:rms}). The broadband noise at low frequency of the PDSs from all the three Insight-HXMT instruments is dominated by a power-law component with index $\sim$1 (\autoref{tab:pds_fit}), which is consistent with the soft state of Cygnus X-1 \citep[e.g. ][]{axelsson_evolution_2005} . The measured fractional rms amplitude above 10 keV is more than two times larger than that below 3 keV (see \autoref{sec:timing}), which is also a characteristic of the soft state of Cygnus X-1 \citep[e.g. ][]{grinberg_long_2014}. Both the X-ray spectral properties of and all-sky monitoring data corresponding to this exposure also show that Cygnus X-1 is in the soft state (\autoref{tab:spec_fit} and \autoref{fig:maxi-bat}). About state classification, it is worth mentioning that Cygnus X-1 has never reached the so-called thermal-dominated or soft state as defined according to BH low-mass X-ray binary phenomenology \citep[BH LMXBs; e.g.][]{remillard_x-ray_2006,belloni_states_2010}. The X-ray spectra of the soft states of other BH LMXBs are dominated by a thermal disk component, while in Cygnus X-1 strong Comptonization and reflection components are also significant \citep{walton_soft_2016,kawano_black_2017,lubinski_distinct_2020}. Furthermore, the fractional rms amplitude measured from soft-state PDSs is usually $\sim$ 20--30 \% in Cygnus X-1 \citep[e.g. ][]{axelsson_probing_2006,grinberg_long_2014}, much larger than the one measured in BH LMXBs in the thermal-dominated state. Keeping in mind these differences, we will refer to this state of Cygnus X-1 simply as the soft state.

 Low-frequency QPOs are quite common in the hard state of BH XRBs. In particular, type-C QPOs are observed with fractional rms amplitude up to 20\% and quality factor $Q$ ($\nu_0$/FWHM) $\geq$10 \citep{casella_abc_2005,motta_quasi_2016}. Type-C QPOs are sometimes also detected in the soft state (e.g. XTE J1550$-564$, GRO J1655$-40$), but with a higher characteristic frequency ($\geq$ 20 Hz) and smaller rms amplitude than those seen in the hard state \citep{homan_correlated_2001,motta_discovery_2012}. Type-C QPOs with similar characteristics are systematically observed in most of the known BH LMXBs. However, type-C QPOs have never been detected in Cygnus X-1 observations \citep{belloni_states_2010,ingram_review_2019}. \citet{pottschmidt_long_2003} has detected weak narrow QPOs ($\nu_{0}\sim$ 0.1--1 Hz) sometimes during the soft state. \citet{rapisarda_modelling_2017} also has detected a millihertz QPO ($\nu_{0}\sim$ 60 mHz) during the soft state. The millihertz QPO we detected has a large quality factor Q ($>13$, see $\nu_{0}$ and FWHM in  \autoref{tab:pds_fit}), which is similar to the type-C QPOs. However, its frequency and rms are smaller than most type-C QPOs observed in other BH XRBs \citep[e.g. ][]{motta_geometrical_2015}. On the other hand, the rms of type-C QPOs usually increases with photon energy and becomes constant above tens of keV \citep[e.g. ][]{casella_study_2004,huang_insight-hxmt_2018}, while the rms of the millihertz QPO we detected in Cygnus X-1 does not show an energy dependence above 3 keV (see \autoref{fig:rms} and \autoref{tab:pds_fit}). Another important difference between type-C QPOs and the millihertz QPO we detected is that the former can be observed for relatively long periods of time ($\sim$ days to tens of days), while the latter appears only in the first 12 ks of this observation. These fundamental differences suggest that the physical mechanism responsible for the observed millihertz QPO in Cygnus X-1 is different from the one producing type-C QPOs in BH LMXBs. 
 
The origin of millihertz QPOs ($<$100 mHz) are still not well understood, although they have been detected in different accreting systems, such as accretion-powered X-ray pulsars \citep[e.g.][]{Psaltis2006} and even ultraluminous X-ray pulsars \citep[e.g.][]{Feng2010}, ultraluminous X-ray sources \citep[e.g.][]{Atapin2019} and BH XRBs \citep[see Table 2 in][and references therein ]{cheng_phase-resolved_2019}. In some BH LMXBs (e.g., H1743$-$322), millihertz QPOs and type-C QPOs are sometimes detected simultaneously and their characteristics mostly appear to be uncorrelated with each other \citep{altamirano_low-frequency_2012}. This supports the conclusion that in BH LMXBs the mechanism producing the millihertz QPO is not related to that producing the type-C QPO. Some of the detected millihertz QPOs appear in high-inclination systems and are thought to have similar origin as to the 1 Hz QPO in dipping neutron star XRBs \citep[e.g.][]{altamirano_low-frequency_2012,armas_padilla_x-ray_2014}. However, the inclination angle of the Cygnus X-1 accretion disk is estimated to be $\sim$ 27$^{\circ}$\citep{orosz_mass_2011}, making unlikely the relation between our detected millihertz QPO and those observed in high-inclination systems. We have noticed that the millihertz QPOs are all detected during the hard state in BH LMXBs, while they are detected during the soft state of high mass X-ray binaries \citep[HMXB, ][]{cheng_phase-resolved_2019},  which indicates different origins in the HMXBs and LMXBs.

The millihertz QPOs in soft states have been previously observed in two other persistent BH HMXBs, namely LMC X-1 \citep{ebisawa_discovery_1989,alam_millihertz_2014} and Cyg X-3 \citep{van_der_klis_transient_1985,koljonen_re-occurrence_2011,pahari_detection_2017}. The millihertz QPOs in Cyg X-3 are thought to be related to major radio flare events \citep{koljonen_re-occurrence_2011}, which are quite unique and have not been observed in other BH XRBs, including Cygnus X-1. LMC X-1 share many common spectral and timing properties during the soft state with Cygnus X-1, even though the X-ray Eddington luminosity of Cygnus X-1 is smaller than that of LMC X-1 \citep[e.g.][]{ruhlen_nature_2011} and LMC X-1 never showed transitions to the hard state. The millihertz QPO in LMC X-1 is also occasionally detected in the soft-state observations, and the frequency also varies \citep{ebisawa_discovery_1989,alam_millihertz_2014}. The rare occurrence in Cygnus X-1, as well as LMC X-1, favors potential mechanisms that would only happen by chance due to some local inhomogeneity in a system embedded in a wind-accretion environment.

It has been proposed that the millihertz QPOs observed in LMC X-1 are due to global disk oscillations \citet{titarchuk_global_2000}, i.e. vertical (normal to the disk) disk oscillations triggered by the gravitational force of the central BH. The frequency of such oscillations can be estimated as
\begin{equation}
\label{eq:f}
    f_{0}\approx2\Big(\frac{R_\mathrm{in}}{6R_\mathrm{g}}\Big)^{-\frac{8}{15}}\Big(\frac{M_\mathrm{BH}}{M_{\sun}}\Big)^{-\frac{8}{15}}\Big(\frac{P_\mathrm{orb}}{3hr}\Big)^{-\frac{7}{15}}\Big(\frac{R_\mathrm{adj}}{R_\mathrm{in}}\Big)^{-0.3} ~ Hz
\end{equation}
where $R_\mathrm{in}$ is the inner radius of the accretion disk and $R_\mathrm{adj}$ is an adjustment radius in the disk, usually of the order of $\sim 2$--$3 R_\mathrm{in}$ for Shakura--Sunyaev geometrically thin disks \citep{shakura_black_1973,titarchuk_mechanisms_1998}. Using $R_\mathrm{adj}=2.5R_\mathrm{in}$, $P_\mathrm{orb}=5.6$ days, $R_\mathrm{in}=2.2R_\mathrm{g}$ (see \autoref{tab:spec_fit}), and $M_\mathrm{BH}=21.2M_{\sun}$ \citep{miller-jones_cygnus_2021}, we obtain an oscillation frequency of $\sim$ 86 mHz, roughly consistent with the frequency of the millihertz QPO we detected. In general, the frequency of the global oscillation depends on the size of the disk and, therefore, on both its inner and outer radius. In \autoref{eq:f}, the disk is assumed to be a geometrically thin Shakura--Sunyaev disk \citep{shakura_black_1973} and the outer disk radius is assumed to be half of the Roche lobe radius \citep{titarchuk_global_2000}. In this scenario, the detection of similar millihertz QPOs at different frequencies in the same and other sources  \citep{rapisarda_modelling_2017,alam_millihertz_2014} could be caused by the variation in the disk size. 

However, as global disk oscillation is the vertical oscillation of the whole disk, the consequently observed modulation is expected to be from the disk emission, where it is mainly in the soft X-ray band (see \autoref{fig:SED}). On the other hand, global disk oscillations would be seen in other BH XRBs not limited to wind accreting systems. The fractional rms amplitude of the detected QPO persisted at a similar amplitude level at least up to 50 keV (see \autoref{fig:rms}), indicating that the QPO either has a hard X-ray spectrum or that the QPO originates primarily in the Comptonization component, such as certain oscillations occur in the disk corona above the standard accretion disk.

\acknowledgments
We thank Chichuan Jin and Joern Wilms for the helpful discussions. This research has made use of MAXI data provided by RIKEN, JAXA, and the MAXI team. This work was supported by part by the Natural Science Foundation of China (grants 11773055, U1838203, and U1938114),  Z.Y. was also supported by the Youth Innovation Promotion Association of CAS (ids. 2020265). R.S. acknowledges the support of the PIFI fellowship of CAS under the project No. 2019PM0016 and China's Postdoctoral International Exchange Program. W.Y. would like to acknowledge partial support by the National Program on Key Research and  Development Project (grant No. 2016YFA0400804).

%

\facilities{Insight-HXMT}


\software{astropy \citep{astropy_collaboration_astropy:_2013},  
          stingray \citep{huppenkothen_stingray_2019}, \texttt{XSPEC}  \citep{arnaud_xspec_1996}
          }





\vspace{50mm}
\bibliography{CygX-1.bbl}{}

\begin{thebibliography}{}
\expandafter\ifx\csname natexlab\endcsname\relax\def\natexlab#1{#1}\fi
\providecommand{\url}[1]{\href{#1}{#1}}
\providecommand{\dodoi}[1]{doi:~\href{http://doi.org/#1}{\nolinkurl{#1}}}
\providecommand{\doeprint}[1]{\href{http://ascl.net/#1}{\nolinkurl{http://ascl.net/#1}}}
\providecommand{\doarXiv}[1]{\href{https://arxiv.org/abs/#1}{\nolinkurl{https://arxiv.org/abs/#1}}}

\bibitem[{Alam {et~al.}(2014)Alam, Dewangan, Belloni, Mukherjee, \&
  Jhingan}]{alam_millihertz_2014}
Alam, M.~S., Dewangan, G.~C., Belloni, T., Mukherjee, D., \& Jhingan, S. 2014,
  Monthly Notices of the Royal Astronomical Society, 445, 4259,
  \dodoi{10.1093/mnras/stu2048}

\bibitem[{Albert {et~al.}(2007)Albert, Aliu, Anderhub, Antoranz, Armada,
  Baixeras, Barrio, Bartko, Bastieri, Becker, Bednarek, Berger, Bigongiari,
  Biland, Bock, Bordas, Bosch-Ramon, Bretz, Britvitch, Camara, Carmona,
  Chilingarian, Coarasa, Commichau, Contreras, Cortina, Costado, Curtef,
  Danielyan, Dazzi, De~Angelis, Delgado, de~los Reyes, De~Lotto,
  Domingo-Santamaría, Dorner, Doro, Errando, Fagiolini, Ferenc, Fernández,
  Firpo, Flix, Fonseca, Font, Fuchs, Galante, García-López, Garczarczyk,
  Gaug, Giller, Goebel, Hakobyan, Hayashida, Hengstebeck, Herrero, Höhne,
  Hose, Hsu, Jacon, Jogler, Kosyra, Kranich, Kritzer, Laille, Lindfors,
  Lombardi, Longo, López, López, Lorenz, Majumdar, Maneva, Mannheim,
  Mansutti, Mariotti, Martínez, Mazin, Merck, Meucci, Meyer, Miranda,
  Mirzoyan, Mizobuchi, Moralejo, Nieto, Nilsson, Ninkovic, Oña-Wilhelmi, Otte,
  Oya, Panniello, Paoletti, Paredes, Pasanen, Pascoli, Pauss, Pegna, Persic,
  Peruzzo, Piccioli, Prandini, Puchades, Raymers, Rhode, Ribó, Rico, Rissi,
  Robert, Rügamer, Saggion, Saito, Sánchez, Sartori, Scalzotto, Scapin,
  Schmitt, Schweizer, Shayduk, Shinozaki, Shore, Sidro, Sillanpää,
  Sobczynska, Stamerra, Stark, Takalo, Temnikov, Tescaro, Teshima, Torres,
  Turini, Vankov, Vitale, Wagner, Wibig, Wittek, Zandanel, Zanin, \&
  Zapatero}]{albert_very_2007}
Albert, J., Aliu, E., Anderhub, H., {et~al.} 2007, The Astrophysical Journal
  Letters, 665, L51, \dodoi{10.1086/521145}

\bibitem[{Altamirano \& Strohmayer(2012)}]{altamirano_low-frequency_2012}
Altamirano, D., \& Strohmayer, T. 2012, The Astrophysical Journal, 754, L23,
  \dodoi{10.1088/2041-8205/754/2/L23}

\bibitem[Atapin et al.(2019)]{Atapin2019} Atapin, K., Fabrika, S., \& Caballero-Garc{\'\i}a, M.~D.\ 2019, \mnras, 486, 2766. doi:10.1093/mnras/stz1027

  
\bibitem[{Armas~Padilla {et~al.}(2014)Armas~Padilla, Wijnands, Altamirano,
  Méndez, Miller, \& Degenaar}]{armas_padilla_x-ray_2014}
Armas~Padilla, M., Wijnands, R., Altamirano, D., {et~al.} 2014, Monthly Notices
  of the Royal Astronomical Society, 439, 3908, \dodoi{10.1093/mnras/stu243}

\bibitem[{Arnaud(1996)}]{arnaud_xspec_1996}
Arnaud, K.~A. 1996, in Astronomical {Data} {Analysis} {Software} and {Systems}
  {V}, Vol. 101, 17.
\newblock \url{http://adsabs.harvard.edu/abs/1996ASPC..101...17A}

\bibitem[{{Astropy Collaboration} {et~al.}(2013){Astropy Collaboration},
  Robitaille, Tollerud, Greenfield, Droettboom, Bray, Aldcroft, Davis,
  Ginsburg, Price-Whelan, Kerzendorf, Conley, Crighton, Barbary, Muna,
  Ferguson, Grollier, Parikh, Nair, Unther, Deil, Woillez, Conseil, Kramer,
  Turner, Singer, Fox, Weaver, Zabalza, Edwards, Azalee~Bostroem, Burke, Casey,
  Crawford, Dencheva, Ely, Jenness, Labrie, Lim, Pierfederici, Pontzen, Ptak,
  Refsdal, Servillat, \& Streicher}]{astropy_collaboration_astropy:_2013}
{Astropy Collaboration}, Robitaille, T.~P., Tollerud, E.~J., {et~al.} 2013,
  Astronomy and Astrophysics, 558, A33, \dodoi{10.1051/0004-6361/201322068}

\bibitem[{Axelsson {et~al.}(2005)Axelsson, Borgonovo, \&
  Larsson}]{axelsson_evolution_2005}
Axelsson, M., Borgonovo, L., \& Larsson, S. 2005, Astronomy and Astrophysics,
  438, 999, \dodoi{10.1051/0004-6361:20042362}

\bibitem[{Axelsson {et~al.}(2006)Axelsson, Borgonovo, \&
  Larsson}]{axelsson_probing_2006}
---. 2006, Astronomy and Astrophysics, 452, 975,
  \dodoi{10.1051/0004-6361:20054397}

\bibitem[{{Belloni} \& {Hasinger}(1990)}]{BelloniHasinger1990}
{Belloni}, T., \& {Hasinger}, G. 1990, \aap, 227, L33

\bibitem[{Belloni {et~al.}(2002)Belloni, Psaltis, \& van~der
  Klis}]{belloni_unified_2002}
Belloni, T., Psaltis, D., \& van~der Klis, M. 2002, The Astrophysical Journal,
  572, 392, \dodoi{10.1086/340290}

\bibitem[{Belloni(2010)}]{belloni_states_2010}
Belloni, T.~M. 2010, in The {Jet} {Paradigm} - {From} {Microquasars} to
  {Quasars}, Vol. 794, eprint: arXiv:0909.2474, 53,
  \dodoi{10.1007/978-3-540-76937-8_3}

\bibitem[{Belloni \& Stella(2014)}]{belloni_fast_2014}
Belloni, T.~M., \& Stella, L. 2014, Space Science Reviews, 183, 43,
  \dodoi{10.1007/s11214-014-0076-0}

\bibitem[{Bolton(1972)}]{bolton_identification_1972}
Bolton, C.~T. 1972, Nature, 235, 271, \dodoi{10.1038/235271b0}

\bibitem[{Bowyer {et~al.}(1965)Bowyer, Byram, Chubb, \& Friedman}]{Bowyer1965}
Bowyer, S., Byram, E.~T., Chubb, T.~A., \& Friedman, H. 1965, Science, 147,
  394, \dodoi{10.1126/science.147.3656.394}

\bibitem[{Casella {et~al.}(2004)Casella, Belloni, Homan, \&
  Stella}]{casella_study_2004}
Casella, P., Belloni, T., Homan, J., \& Stella, L. 2004, Astronomy and
  Astrophysics, 426, 587, \dodoi{10.1051/0004-6361:20041231}

\bibitem[{Casella {et~al.}(2005)Casella, Belloni, \& Stella}]{casella_abc_2005}
Casella, P., Belloni, T., \& Stella, L. 2005, The Astrophysical Journal, 629,
  403, \dodoi{10.1086/431174}

\bibitem[{Cheng {et~al.}(2019)Cheng, Méndez, Altamirano, Beri, \&
  Wang}]{cheng_phase-resolved_2019}
Cheng, Z., Méndez, M., Altamirano, D., Beri, A., \& Wang, Y. 2019, Monthly
  Notices of the Royal Astronomical Society, 482, 550,
  \dodoi{10.1093/mnras/sty2695}

\bibitem[{Cui {et~al.}(1997{\natexlab{a}})Cui, Heindl, Rothschild, Zhang,
  Jahoda, \& Focke}]{cui_rossi_1997}
Cui, W., Heindl, W.~A., Rothschild, R.~E., {et~al.} 1997{\natexlab{a}}, The
  Astrophysical Journal Letters, 474, L57, \dodoi{10.1086/310419}

\bibitem[{Cui {et~al.}(1997{\natexlab{b}})Cui, Zhang, Focke, \&
  Swank}]{cui_temporal_1997}
Cui, W., Zhang, S.~N., Focke, W., \& Swank, J.~H. 1997{\natexlab{b}}, The
  Astrophysical Journal, 484, 383, \dodoi{10.1086/304341}

\bibitem[{Done {et~al.}(2007)Done, Gierliński, \&
  Kubota}]{done_modelling_2007}
Done, C., Gierliński, M., \& Kubota, A. 2007, Astronomy and Astrophysics
  Review, 15, 1, \dodoi{10.1007/s00159-007-0006-1}

\bibitem[{Ebisawa {et~al.}(1989)Ebisawa, Mitsuda, \&
  Inoue}]{ebisawa_discovery_1989}
Ebisawa, K., Mitsuda, K., \& Inoue, H. 1989, Publications of the Astronomical
  Society of Japan, 41, 519.
\newblock \url{http://adsabs.harvard.edu/abs/1989PASJ...41..519E}

\bibitem[{Fabian {et~al.}(2012)Fabian, Wilkins, Miller, Reis, Reynolds,
  Cackett, Nowak, Pooley, Pottschmidt, Sanders, Ross, \&
  Wilms}]{fabian_determination_2012}
Fabian, A.~C., Wilkins, D.~R., Miller, J.~M., {et~al.} 2012, Monthly Notices of
  the Royal Astronomical Society, 424, 217,
  \dodoi{10.1111/j.1365-2966.2012.21185.x}

\bibitem[Feng et al.(2010)]{Feng2010} Feng, H., Rao, F., \& Kaaret, P.\ 2010, \apjl, 710, L137. doi:10.1088/2041-8205/710/2/L137
  
\bibitem[{Foreman-Mackey {et~al.}(2013)Foreman-Mackey, Hogg, Lang, \&
  Goodman}]{foreman-mackey_emcee:_2013}
Foreman-Mackey, D., Hogg, D.~W., Lang, D., \& Goodman, J. 2013, Publications of
  the Astronomical Society of the Pacific, 125, 306, \dodoi{10.1086/670067}

\bibitem[{Gallo {et~al.}(2005)Gallo, Fender, Kaiser, Russell, Morganti,
  Oosterloo, \& Heinz}]{gallo_dark_2005}
Gallo, E., Fender, R., Kaiser, C., {et~al.} 2005, Nature, 436, 819,
  \dodoi{10.1038/nature03879}

\bibitem[{Gierliński {et~al.}(1999)Gierliński, Zdziarski, Poutanen, Coppi,
  Ebisawa, \& Johnson}]{gierlinski_radiation_1999}
Gierliński, M., Zdziarski, A.~A., Poutanen, J., {et~al.} 1999, Monthly Notices
  of the Royal Astronomical Society, 309, 496,
  \dodoi{10.1046/j.1365-8711.1999.02875.x}

\bibitem[{Grinberg {et~al.}(2013)Grinberg, Hell, Pottschmidt, Böck, Nowak,
  Rodriguez, Bodaghee, Cadolle~Bel, Case, Hanke, Kühnel, Markoff, Pooley,
  Rothschild, Tomsick, Wilson-Hodge, \& Wilms}]{grinberg_long_2013}
Grinberg, V., Hell, N., Pottschmidt, K., {et~al.} 2013, Astronomy and
  Astrophysics, 554, A88, \dodoi{10.1051/0004-6361/201321128}

\bibitem[{Grinberg {et~al.}(2014)Grinberg, Pottschmidt, Böck, Schmid, Nowak,
  Uttley, Tomsick, Rodriguez, Hell, Markowitz, Bodaghee, Cadolle~Bel,
  Rothschild, \& Wilms}]{grinberg_long_2014}
Grinberg, V., Pottschmidt, K., Böck, M., {et~al.} 2014, Astronomy and
  Astrophysics, 565, A1, \dodoi{10.1051/0004-6361/201322969}

\bibitem[{Herrero {et~al.}(1995)Herrero, Kudritzki, Gabler, Vilchez, \&
  Gabler}]{herrero_fundamental_1995}
Herrero, A., Kudritzki, R.~P., Gabler, R., Vilchez, J.~M., \& Gabler, A. 1995,
  Astronomy and Astrophysics, 297, 556.
\newblock \url{http://adsabs.harvard.edu/abs/1995A%26A...297..556H}

\bibitem[{Homan {et~al.}(2001)Homan, Wijnands, van~der Klis, Belloni, van
  Paradijs, Klein-Wolt, Fender, \& Méndez}]{homan_correlated_2001}
Homan, J., Wijnands, R., van~der Klis, M., {et~al.} 2001, The Astrophysical
  Journal Supplement Series, 132, 377, \dodoi{10.1086/318954}

\bibitem[{Huang {et~al.}(2018)Huang, Qu, Zhang, Bu, Chen, Tao, Zhang, Lu, Li,
  Song, Xu, Cao, Chen, Liu, Chang, Yu, Weng, Hou, Kong, Xie, Zhang, Zhou,
  Chang, Chen, Chen, Chen, Chen, Cui, Cui, Deng, Dong, Du, Fu, Gao, Gao, Gao,
  Ge, Gu, Guan, Gungor, Guo, Han, Hu, Huo, Ji, Jia, Jiang, Jiang, Jin, Jin, Li,
  Li, Li, Li, Li, Li, Li, Li, Li, Li, Li, Liang, Liao, Liu, Liu, Liu, Liu, Liu,
  Liu, Lu, Lu, Luo, Ma, Meng, Nang, Nie, Ou, Sai, Shang, Sun, Tan, Tao, Tuo,
  Wang, Wang, Wang, Wang, Wang, Wen, Wu, Wu, Xiao, Xiong, Xu, Yan, Yang, Yang,
  Yang, Zhang, Zhang, Zhang, Zhang, Zhang, Zhang, Zhang, Zhang, Zhang, Zhang,
  Zhang, Zhang, Zhang, Zhang, Zhang, Zhang, Zhang, Zhang, Zhao, Zhao, Zhao,
  Zheng, Zhu, Zhu, Zou, \& Collaboration}]{huang_insight-hxmt_2018}
Huang, Y., Qu, J.~L., Zhang, S.~N., {et~al.} 2018, The Astrophysical Journal,
  866, 122, \dodoi{10.3847/1538-4357/aade4c}

\bibitem[{Huppenkothen {et~al.}(2017)Huppenkothen, Younes, Ingram, Kouveliotou,
  Göğüş, Bachetti, Sánchez-Fernández, Chenevez, Motta, van~der Klis,
  Granot, Gehrels, Kuulkers, Tomsick, \& Walton}]{huppenkothen_detection_2017}
Huppenkothen, D., Younes, G., Ingram, A., {et~al.} 2017, The Astrophysical
  Journal, 834, 90, \dodoi{10.3847/1538-4357/834/1/90}

\bibitem[{Huppenkothen {et~al.}(2019)Huppenkothen, Bachetti, Stevens, Migliari,
  Balm, Hammad, Khan, Mishra, Rashid, Sharma, Martinez~Ribeiro, \&
  Valles~Blanco}]{huppenkothen_stingray_2019}
Huppenkothen, D., Bachetti, M., Stevens, A.~L., {et~al.} 2019, The
  Astrophysical Journal, 881, 39, \dodoi{10.3847/1538-4357/ab258d}

\bibitem[{Ingram \& Motta(2019)}]{ingram_review_2019}
Ingram, A.~R., \& Motta, S.~E. 2019, New Astronomy Reviews, 85, 101524,
  \dodoi{10.1016/j.newar.2020.101524}

\bibitem[{Israel \& Stella(1996)}]{israel_new_1996}
Israel, G.~L., \& Stella, L. 1996, The Astrophysical Journal, 468, 369,
  \dodoi{10.1086/177697}

\bibitem[{Kantzas {et~al.}(2021)Kantzas, Markoff, Beuchert, Lucchini, Chhotray,
  Ceccobello, Tetarenko, Miller-Jones, Bremer, Garcia, Grinberg, Uttley, \&
  Wilms}]{kantzas_new_2021}
Kantzas, D., Markoff, S., Beuchert, T., {et~al.} 2021, Monthly Notices of the
  Royal Astronomical Society, 500, 2112, \dodoi{10.1093/mnras/staa3349}

\bibitem[{Kawano {et~al.}(2017)Kawano, Done, Yamada, Takahashi, Axelsson, \&
  Fukazawa}]{kawano_black_2017}
Kawano, T., Done, C., Yamada, S., {et~al.} 2017, Publications of the
  Astronomical Society of Japan, 69, 36, \dodoi{10.1093/pasj/psx009}

\bibitem[{Koljonen {et~al.}(2011)Koljonen, Hannikainen, \&
  McCollough}]{koljonen_re-occurrence_2011}
Koljonen, K. I.~I., Hannikainen, D.~C., \& McCollough, M.~L. 2011, Monthly
  Notices of the Royal Astronomical Society, 416, L84,
  \dodoi{10.1111/j.1745-3933.2011.01104.x}

\bibitem[{Krimm {et~al.}(2013)Krimm, Holland, Corbet, Pearlman, Romano, Kennea,
  Bloom, Barthelmy, Baumgartner, Cummings, Gehrels, Lien, Markwardt, Palmer,
  Sakamoto, Stamatikos, \& Ukwatta}]{krimm_swiftbat_2013}
Krimm, H.~A., Holland, S.~T., Corbet, R. H.~D., {et~al.} 2013, The
  Astrophysical Journal Supplement Series, 209, 14,
  \dodoi{10.1088/0067-0049/209/1/14}

\bibitem[{Leahy {et~al.}(1983)Leahy, Elsner, \&
  Weisskopf}]{leahy_searches_1983}
Leahy, D.~A., Elsner, R.~F., \& Weisskopf, M.~C. 1983, The Astrophysical
  Journal, 272, 256, \dodoi{10.1086/161288}

\bibitem[{Li {et~al.}(2020)Li, Li, Tan, Yang, Ge, Zhang, Tuo, Wu, Liao, Zhang,
  Song, Zhang, Qu, Zhang, Lu, Xu, Liu, Cao, Chen, Nie, Zhao, \&
  Li}]{li_-flight_2020}
Li, X., Li, X., Tan, Y., {et~al.} 2020, Journal of High Energy Astrophysics,
  27, 64, \dodoi{10.1016/j.jheap.2020.02.009}

\bibitem[{Lubiński {et~al.}(2020)Lubiński, Filothodoros, Zdziarski, \&
  Pooley}]{lubinski_distinct_2020}
Lubiński, P., Filothodoros, A., Zdziarski, A.~A., \& Pooley, G. 2020, The
  Astrophysical Journal, 896, 101, \dodoi{10.3847/1538-4357/ab9311}

\bibitem[{Matsuoka {et~al.}(2009)Matsuoka, Kawasaki, Ueno, Tomida, Kohama,
  Suzuki, Adachi, Ishikawa, Mihara, Sugizaki, Isobe, Nakagawa, Tsunemi, Miyata,
  Kawai, Kataoka, Morii, Yoshida, Negoro, Nakajima, Ueda, Chujo, Yamaoka,
  Yamazaki, Nakahira, You, Ishiwata, Miyoshi, Eguchi, Hiroi, Katayama, \&
  Ebisawa}]{matsuoka_maxi_2009}
Matsuoka, M., Kawasaki, K., Ueno, S., {et~al.} 2009, Publications of the
  Astronomical Society of Japan, 61, 999, \dodoi{10.1093/pasj/61.5.999}

\bibitem[{Miller-Jones {et~al.}(2021)Miller-Jones, Bahramian, Orosz, Mandel,
  Gou, Maccarone, Neijssel, Zhao, Ziółkowski, Reid, Uttley, Zheng, Byun,
  Dodson, Grinberg, Jung, Kim, Marcote, Markoff, Rioja, Rushton, Russell,
  Sivakoff, Tetarenko, Tudose, \& Wilms}]{miller-jones_cygnus_2021}
Miller-Jones, J. C.~A., Bahramian, A., Orosz, J.~A., {et~al.} 2021, arXiv
  e-prints, 2102, arXiv:2102.09091.
\newblock \url{http://adsabs.harvard.edu/abs/2021arXiv210209091M}

\bibitem[{Motta {et~al.}(2012)Motta, Homan, Muñoz~Darias, Casella, Belloni,
  Hiemstra, \& Méndez}]{motta_discovery_2012}
Motta, S., Homan, J., Muñoz~Darias, T., {et~al.} 2012, Monthly Notices of the
  Royal Astronomical Society, 427, 595,
  \dodoi{10.1111/j.1365-2966.2012.22037.x}

\bibitem[{Motta(2016)}]{motta_quasi_2016}
Motta, S.~E. 2016, Astronomische Nachrichten, 337, 398,
  \dodoi{10.1002/asna.201612320}

\bibitem[{Motta {et~al.}(2015)Motta, Casella, Henze, Muñoz-Darias, Sanna,
  Fender, \& Belloni}]{motta_geometrical_2015}
Motta, S.~E., Casella, P., Henze, M., {et~al.} 2015, Monthly Notices of the
  Royal Astronomical Society, 447, 2059, \dodoi{10.1093/mnras/stu2579}

\bibitem[{Niedźwiecki {et~al.}(2019)Niedźwiecki, Szanecki, \&
  Zdziarski}]{niedzwiecki_improved_2019}
Niedźwiecki, A., Szanecki, M., \& Zdziarski, A.~A. 2019, Monthly Notices of
  the Royal Astronomical Society, 485, 2942, \dodoi{10.1093/mnras/stz487}

\bibitem[{Nowak {et~al.}(1999)Nowak, Vaughan, Wilms, Dove, \&
  Begelman}]{nowak_rossi_1999}
Nowak, M.~A., Vaughan, B.~A., Wilms, J., Dove, J.~B., \& Begelman, M.~C. 1999,
  The Astrophysical Journal, 510, 874, \dodoi{10.1086/306610}

\bibitem[{Orosz {et~al.}(2011)Orosz, McClintock, Aufdenberg, Remillard, Reid,
  Narayan, \& Gou}]{orosz_mass_2011}
Orosz, J.~A., McClintock, J.~E., Aufdenberg, J.~P., {et~al.} 2011, The
  Astrophysical Journal, 742, 84, \dodoi{10.1088/0004-637X/742/2/84}

\bibitem[{Pahari {et~al.}(2017)Pahari, McHardy, Mallick, Dewangan, \&
  Misra}]{pahari_detection_2017}
Pahari, M., McHardy, I.~M., Mallick, L., Dewangan, G.~C., \& Misra, R. 2017,
  Monthly Notices of the Royal Astronomical Society, 470, 3239,
  \dodoi{10.1093/mnras/stx1455}

\bibitem[Psaltis(2006)]{Psaltis2006} Psaltis, D.\ 2006, Compact stellar X-ray sources, 1  

\bibitem[{Paul {et~al.}(1998)Paul, Agrawal, \& Rao}]{paul_low_1998}
Paul, B., Agrawal, P.~C., \& Rao, A.~R. 1998, Journal of Astrophysics and
  Astronomy, 19, 55, \dodoi{10.1007/BF02714891}

\bibitem[{Pottschmidt {et~al.}(2003)Pottschmidt, Wilms, Nowak, Pooley,
  Gleissner, Heindl, Smith, Remillard, \& Staubert}]{pottschmidt_long_2003}
Pottschmidt, K., Wilms, J., Nowak, M.~A., {et~al.} 2003, Astronomy and
  Astrophysics, 407, 1039, \dodoi{10.1051/0004-6361:20030906}

\bibitem[{Poutanen \& Svensson(1996)}]{poutanen_two-phase_1996}
Poutanen, J., \& Svensson, R. 1996, The Astrophysical Journal, 470, 249,
  \dodoi{10.1086/177865}

\bibitem[{Rapisarda {et~al.}(2017)Rapisarda, Ingram, \& van~der
  Klis}]{rapisarda_modelling_2017}
Rapisarda, S., Ingram, A., \& van~der Klis, M. 2017, Monthly Notices of the
  Royal Astronomical Society, 472, 3821, \dodoi{10.1093/mnras/stx2110}

\bibitem[{Remillard \& McClintock(2006)}]{remillard_x-ray_2006}
Remillard, R.~A., \& McClintock, J.~E. 2006, Annual Review of Astronomy and
  Astrophysics, 44, 49, \dodoi{10.1146/annurev.astro.44.051905.092532}

\bibitem[{Ruhlen {et~al.}(2011)Ruhlen, Smith, \& Swank}]{ruhlen_nature_2011}
Ruhlen, L., Smith, D.~M., \& Swank, J.~H. 2011, The Astrophysical Journal, 742,
  75, \dodoi{10.1088/0004-637X/742/2/75}

\bibitem[{Shakura \& Sunyaev(1973)}]{shakura_black_1973}
Shakura, N.~I., \& Sunyaev, R.~A. 1973, Astronomy and Astrophysics, 24, 337.
\newblock \url{http://adsabs.harvard.edu/abs/1973A%26A....24..337S}

\bibitem[{Shaposhnikov \& Titarchuk(2006)}]{shaposhnikov_comprehensive_2006}
Shaposhnikov, N., \& Titarchuk, L. 2006, The Astrophysical Journal, 643, 1098,
  \dodoi{10.1086/503272}

\bibitem[{Tananbaum {et~al.}(1972)Tananbaum, Gursky, Kellogg, Giacconi, \&
  Jones}]{tananbaum_observation_1972}
Tananbaum, H., Gursky, H., Kellogg, E., Giacconi, R., \& Jones, C. 1972, The
  Astrophysical Journal Letters, 177, L5, \dodoi{10.1086/181042}

\bibitem[{Titarchuk {et~al.}(1998)Titarchuk, Lapidus, \&
  Muslimov}]{titarchuk_mechanisms_1998}
Titarchuk, L., Lapidus, I., \& Muslimov, A. 1998, The Astrophysical Journal,
  499, 315, \dodoi{10.1086/305642}

\bibitem[{Titarchuk \& Osherovich(2000)}]{titarchuk_global_2000}
Titarchuk, L., \& Osherovich, V. 2000, The Astrophysical Journal Letters, 542,
  L111, \dodoi{10.1086/312935}

\bibitem[{Tomsick {et~al.}(2014)Tomsick, Nowak, Parker, Miller, Fabian,
  Harrison, Bachetti, Barret, Boggs, Christensen, Craig, Forster, Fürst,
  Grefenstette, Hailey, King, Madsen, Natalucci, Pottschmidt, Ross, Stern,
  Walton, Wilms, \& Zhang}]{tomsick_reflection_2014}
Tomsick, J.~A., Nowak, M.~A., Parker, M., {et~al.} 2014, The Astrophysical
  Journal, 780, 78, \dodoi{10.1088/0004-637X/780/1/78}

\bibitem[{van~der Klis(1989)}]{van_der_klis_fourier_1989}
van~der Klis, M. 1989, in Timing {Neutron} {Stars}: proceedings of the {NATO}
  {Advanced} {Study} {Institute} on {Timing} {Neutron} {Stars}, Vol. 262, 27.
\newblock \url{http://adsabs.harvard.edu/abs/1989ASIC..262...27V}

\bibitem[{van~der Klis \& Jansen(1985)}]{van_der_klis_transient_1985}
van~der Klis, M., \& Jansen, F.~A. 1985, Nature, 313, 768,
  \dodoi{10.1038/313768a0}

\bibitem[{Vaughan(2005)}]{vaughan_simple_2005}
Vaughan, S. 2005, Astronomy and Astrophysics, 431, 391,
  \dodoi{10.1051/0004-6361:20041453}

\bibitem[{Vaughan(2010)}]{vaughan_bayesian_2010}
---. 2010, Monthly Notices of the Royal Astronomical Society, 402, 307,
  \dodoi{10.1111/j.1365-2966.2009.15868.x}

\bibitem[{Verner {et~al.}(1996)Verner, Ferland, Korista, \&
  Yakovlev}]{verner_atomic_1996}
Verner, D.~A., Ferland, G.~J., Korista, K.~T., \& Yakovlev, D.~G. 1996, The
  Astrophysical Journal, 465, 487, \dodoi{10.1086/177435}

\bibitem[{Walton {et~al.}(2016)Walton, Tomsick, Madsen, Grinberg, Barret,
  Boggs, Christensen, Clavel, Craig, Fabian, Fuerst, Hailey, Harrison, Miller,
  Parker, Rahoui, Stern, Tao, Wilms, \& Zhang}]{walton_soft_2016}
Walton, D.~J., Tomsick, J.~A., Madsen, K.~K., {et~al.} 2016, The Astrophysical
  Journal, 826, 87, \dodoi{10.3847/0004-637X/826/1/87}

\bibitem[{Webster \& Murdin(1972)}]{webster_cygnus_1972}
Webster, B.~L., \& Murdin, P. 1972, Nature, 235, 37, \dodoi{10.1038/235037a0}

\bibitem[{Wijnands {et~al.}(1999)Wijnands, Homan, \& van~der
  Klis}]{wijnands_complex_1999}
Wijnands, R., Homan, J., \& van~der Klis, M. 1999, The Astrophysical Journal
  Letters, 526, L33, \dodoi{10.1086/312365}

\bibitem[{Wilms {et~al.}(2000)Wilms, Allen, \& McCray}]{wilms_absorption_2000}
Wilms, J., Allen, A., \& McCray, R. 2000, The Astrophysical Journal, 542, 914,
  \dodoi{10.1086/317016}

\bibitem[{Yuan \& Narayan(2014)}]{yuan_hot_2014}
Yuan, F., \& Narayan, R. 2014, Annual Review of Astronomy and Astrophysics, 52,
  529, \dodoi{10.1146/annurev-astro-082812-141003}

\bibitem[{Zdziarski {et~al.}(1996)Zdziarski, Johnson, \&
  Magdziarz}]{zdziarski_broad-band_1996}
Zdziarski, A.~A., Johnson, W.~N., \& Magdziarz, P. 1996, Monthly Notices of the
  Royal Astronomical Society, 283, 193, \dodoi{10.1093/mnras/283.1.193}

\bibitem[{Zhang {et~al.}(2020)Zhang, Li, Lu, Song, Xu, Liu, Chen, Cao, Bu,
  Chang, Chen, Chen, Chen, Chen, Chen, Cui, Cui, Deng, Dong, Du, Fu, Gao, Gao,
  Gao, Ge, Gu, Guan, Gungor, Guo, Han, Hu, Huang, Huo, Jia, Jiang, Jiang, Jin,
  Jin, Li, Li, Li, Li, Li, Li, Li, Li, Li, Li, Li, Liang, Liao, Liu, Liu, Liu,
  Liu, Liu, Liu, Lu, Lu, Luo, Ma, Meng, Nang, Nie, Ou, Qu, Sai, Shang, Shen,
  Sun, Tan, Tao, Tuo, Wang, Wang, Wang, Wang, Wang, Wang, Wang, Wen, Wu, Wu,
  Wu, Xiao, Xiong, Yan, Yang, Yang, Yang, Yi, Yuan, Zhang, Zhang, Zhang, Zhang,
  Zhang, Zhang, Zhang, Zhang, Zhang, Zhang, Zhang, Zhang, Zhang, Zhang, Zhang,
  Zhang, Zhang, Zhang, Zhang, Zhang, Zhao, Zhao, Zheng, Zhou, Zhu, Zhu, Zhuang,
  \& {Insight-HXMT Team}}]{zhang_overview_2020}
Zhang, S.-N., Li, T., Lu, F., {et~al.} 2020, Science China Physics, Mechanics,
  and Astronomy, 63, 249502, \dodoi{10.1007/s11433-019-1432-6}

\bibitem[{Zhao {et~al.}(2021)Zhao, Gou, Dong, Zheng, Steiner, Miller-Jones,
  Bahramian, Orosz, \& Feng}]{zhao_re-estimating_2021}
Zhao, X., Gou, L., Dong, Y., {et~al.} 2021, arXiv e-prints, 2102,
  arXiv:2102.09093.
\newblock \url{http://adsabs.harvard.edu/abs/2021arXiv210209093Z}

\bibitem[{Zhao {et~al.}(2020)Zhao, Dong, Gou, Feng, Jia, Li, Liao, Liu, Zheng,
  Zhang, Qu, Song, Zhang, Bu, Cai, Cao, Chang, Chen, Chen, Chen, Chen, Chen,
  Chen, Cui, Cui, Deng, Dong, Du, Fu, Gao, Gao, Gao, Ge, Gu, Guan, Guo, Han,
  Huang, Huo, Jia, Jiang, Jiang, Jin, Jin, Kong, Li, Li, Li, Li, Li, Li, Li,
  Li, Li, Li, Li, Liang, Liao, Liu, Liu, Liu, Liu, Liu, Liu, Liu, Lu, Lu, Lu,
  Luo, Luo, Ma, Meng, Nang, Nie, Ou, Sai, Shang, Song, Sun, Tan, Tao, Tuo,
  Wang, Wang, Wang, Wang, Wang, Wen, Wu, Wu, Wu, Xiao, Xiao, Xiong, Xu, Yang,
  Yang, Yang, Yang, Yi, Yin, You, Zhang, Zhang, Zhang, Zhang, Zhang, Zhang,
  Zhang, Zhang, Zhang, Zhang, Zhang, Zhang, Zhang, Zhang, Zhang, Zhang, Zhao,
  Zhao, Zheng, Zhou, Zhou, Zhu, Zhu, \& Zhuang}]{zhao_confirming_2020}
Zhao, X.-S., Dong, Y.-T., Gou, L.-J., {et~al.} 2020, Journal of High Energy
  Astrophysics, 27, 53, \dodoi{10.1016/j.jheap.2020.03.001}

\end{thebibliography}



\end{document}